\documentclass[aps,pra,twocolumn,groupedaddress,showpacs]{revtex4-1}

\usepackage{graphicx}

\newcommand{\figurewidth}{\columnwidth}
\begin{document}

\title{Dielectronic recombination of xenonlike tungsten ions}

\author{S. Schippers}
\email[]{Stefan.Schippers@physik.uni-giessen.de}
\homepage[]{http://www.uni-giessen.de/cms/iamp}
\affiliation{Institut f\"{u}r Atom- und Molek\"{u}lphysik, Justus-Liebig-Universit\"{a}t Giessen, Leihgesterner Weg 217, 35392 Giessen, Germany}

\author{D. Bernhardt}
\affiliation{Institut f\"{u}r Atom- und Molek\"{u}lphysik, Justus-Liebig-Universit\"{a}t Giessen, Leihgesterner Weg 217, 35392 Giessen, Germany}

\author{A. M\"{u}ller}
\affiliation{Institut f\"{u}r Atom- und Molek\"{u}lphysik, Justus-Liebig-Universit\"{a}t Giessen, Leihgesterner Weg 217, 35392 Giessen, Germany}

\author{C. Krantz}
\affiliation{Max-Planck-Institut f\"{u}r Kernphysik, Saupfercheckweg 1, 69117 Heidelberg, Germany}

\author{M. Grieser}
\affiliation{Max-Planck-Institut f\"{u}r Kernphysik, Saupfercheckweg 1, 69117 Heidelberg, Germany}

\author{R. Repnow}
\affiliation{Max-Planck-Institut f\"{u}r Kernphysik, Saupfercheckweg 1, 69117 Heidelberg, Germany}

\author{A. Wolf}
\affiliation{Max-Planck-Institut f\"{u}r Kernphysik, Saupfercheckweg 1, 69117 Heidelberg, Germany}

\author{M. Lestinsky}
\affiliation{GSI Helmholtzzentrum f\"{u}r Schwerionenforschung mbH, Planckstra{\ss}e 1, 64291 Darmstadt, Germany}

\author{M. Hahn}
\affiliation{Columbia Astrophysics Laboratory, Columbia University, 550 West 120th Street, New York, NY 10027, USA}

\author{O. Novotn\'y}
\affiliation{Columbia Astrophysics Laboratory, Columbia University, 550 West 120th Street, New York, NY 10027, USA}

\author{D. W. Savin}
\affiliation{Columbia Astrophysics Laboratory, Columbia University, 550 West 120th Street, New York, NY 10027, USA}

\date{\today}

\begin{abstract}
Dielectronic recombination (DR) of xenonlike W$^{20+}$ forming W$^{19+}$ has been studied experimentally at a heavy-ion storage-ring. A merged-beams method has been employed for obtaining absolute rate coefficients for electron-ion recombination in the collision energy range 0--140 eV. The measured rate coefficient is dominated by strong DR resonances even at the lowest experimental energies. At plasma temperatures where the fractional abundance of W$^{20+}$  is expected to peak in a fusion plasma, the experimentally derived plasma recombination rate coefficient is over a factor of 4 larger than the theoretically-calculated rate coefficient which is currently used in fusion plasma modeling. The largest part of this discrepancy stems most probably from the neglect in the theoretical calculations of DR associated with fine-structure excitations of the W$^{20+}$([Kr]$4d^{10}\,4f^8$) ion core.
\end{abstract}

\pacs{34.80.Lx,52.20.Fs}

\maketitle

\section{Introduction\label{sec:intro}}

Atomic spectroscopy and collision processes involving tungsten ions currently receive much attention, since tungsten is used as a wall material in nuclear fusion reactors \cite{Neu2005,Neu2009}. Consequently, tungsten ions are expected to be prominent impurities in fusion plasmas. Radiation from excited tungsten ions leads to substantial  plasma cooling which has to be well controlled in order to maintain the conditions for nuclear fusion.  Thus, a comprehensive knowledge of atomic energy levels and collision cross sections is required for a thorough understanding of the spatial and temporal evolution of the tungsten charge states and emission spectra in fusion plasmas \cite{Peacock2008,Puetterich2008,Puetterich2010}. To date, only a small fraction of the needed atomic data has been derived from experimental measurements and most comes from theory \cite{Reader2009}. Because of the complexity of the atomic structure of most tungsten ions, the theoretical methods require the use of approximations in order to become tractable. The associated uncertainties in the calculated cross sections are generally difficult to assess.

The situation is particularly problematic for electron-ion recombination which is an important process governing the charge balance in plasmas and which also leads to the population of excited states in the recombined ion. For tungsten ions, no experimental benchmarks in the form of absolute rate coefficients are yet available. Full quantum mechanical calculations of rate coefficients for dielectronic recombination (DR) have been carried out only for a very few charge states \footnote{Throughout this paper, ions are identified by their charge state before recombination.}, i.e., for Ne-like W$^{64+}$ \cite{Behar1999,Safronova2009}, Ar-like W$^{56+}$ \cite{Peleg1998}, and Y-like W$^{35+}$ \cite{Ballance2010}. Rate coefficients for radiative recombination (RR) have been calculated for selected closed-shell and hydrogenlike tungsten ions employing the relativistic Dirac-Fock method \cite{Trzhaskovskaya2008,Trzhaskovskaya2010}.

Current plasma modeling \cite{Puetterich2008,Puetterich2010} uses DR rate coefficients from the ADAS data base \cite{adas,Whiteford2004} which are based on the semi-empirical Burgess formula \cite{Burgess1965}. Since the uncertainties of these DR rate coefficients were considered to be larger than those of rate coefficients for other relevant atomic processes, the DR rate  coefficients were multiplied by temperature independent ad-hoc scaling factors in order to bring the model predictions into agreement with measured emission spectra from a tokamak plasma \cite{Puetterich2008}. For W$^{q+}$ ions these scaling factors range from 0.26 to 2.25, depending on charge state~$q$.

To the best of our knowledge no experimental measurements of absolute electron-ion recombination rate coefficients of tungsten ions have been carried out so far. Experiments with highly charged tungsten ions have hitherto mainly focussed on photon emission spectroscopy in the extreme ultraviolet and X-ray spectral ranges from tokamak and stellarator plasmas \cite{Neu1997,Puetterich2008,Harte2010} and  electron-beam ion-traps (EBITs) \cite{Utter2000a,Utter2002,Puetterich2005,Ralchenko2006,Ralchenko2007a,Radtke2007,Watanabe2007,Ralchenko2008b,
Biedermann2009,Biedermann2009a,Podpaly2009,Clementson2010}. Results have been obtained for a number charge states in the range $q$=27--67. These provide vitally needed information for plasma diagnostics. Additionally, cross sections for electron-impact ionization have been measured for  $q$=1--10 \cite{Stenke1995c}, and photoionization measurements using synchrotron radiation have been reported for $q$=1--3, and 5 \cite{Mueller2011}.

Here, we present the absolute experimental rate coefficient of Xe-like W$^{20+}$ recombining to form Cs-like W$^{19+}$. For these measurements an electron-ion merged beams technique \cite{Phaneuf1999} has been employed at a heavy-ion storage ring \cite{Mueller1997c}. Details of the experimental procedures are given in section \ref{sec:exp}. The experimental results are presented and discussed in section \ref{sec:res}, and conclusions are given in section \ref{sec:conc}.

\section{Experiment}\label{sec:exp}

The experiment was performed using the accelerator and storage-ring facilities of the  Max-Planck-Institut f\"{u}r Kernphysik (MPIK) in Heidelberg, Germany. Negatively-charged tungsten ions were produced from tungsten carbide in a cesium sputter ion-source and injected into a tandem accelerator. Positively charged W$^{20+}$ ions were obtained by passing the ion beam twice (at different energies) through thin carbon foils and by appropriate charge-to-mass selection in a dipole magnet. The time-averaged electrical current of  $^{186}$W$^{20+}$ ions after this magnet was typically 100~pA at an ion energy of 192~MeV.  After multiturn injection of five ion pulses into the storage-ring TSR, circulating ion currents of the order of a few nA were expected considering the stacking performance of previous experiments \cite{Grieser1991}. The velocity of the stored ions on their closed orbit was 4.7\% of the speed of light. Electron cooling \cite{Poth1990} was applied to reduce the momentum spread and diameter of the ion beam. Measurements started 2~s after injection to allow all excited states with expected lifetimes of up to several hundred milliseconds to decay to the W$^{20+}$ ground configuration (see discussion below).

The TSR storage ring is equipped with two electron-beam arrangements, dubbed `Cooler' and `Target' \cite{Wolf2006c}, which can be used for electron cooling and for electron-ion collision studies. For the present experiment the Target was used for electron cooling and the Cooler served as an electron target for the electron-ion recombination measurements. This choice was made since the Cooler featured a higher electron density than the Target resulting in larger product-ion count rates. Recombined W$^{19+}$ ions were separated from the primary W$^{20+}$ beam in the first bending dipole magnet behind the Cooler and counted with a single particle detector \cite{Rinn1982} with effectively 100\% efficiency. Data were taken at different electron-ion collision energies. These were adjusted by setting the Cooler cathode voltage appropriately \cite{Kilgus1992}.

Usually absolute merged-beams recombination rate coefficients are derived from the measured recombination count rates by an appropriate normalization to the electron and ion currents \cite{Kilgus1992,Schmidt2007b}. In the present experiment the ion current was so low that it could not be properly measured. Relative recombination rate coefficients were determined by normalizing to a proxy of the ion current. Here we used the recombination signal at a fixed non-zero electron-ion collision energy $E_\mathrm{ref}$. The reference energy $E_\mathrm{ref}=131.8$~eV was chosen such that the recombination count rate at this electron-ion collision energy was predominantly due to electron-capture during collisions with residual gas particles. The reference measurements were interleaved with the measurements at energies $E_\nu$ ($\nu=1,2,3,\ldots$) such that a sequence of electron-ion collision energies was stepped through during one measurement cycle (i.e., $E_1 - E_\mathrm{ref} - E_2 - E_\mathrm{ref} -E_3-E_\mathrm{ref}-\ldots$). The dwell time at each energy was 20~ms. Only the last 10~ms were used for data taking in order to allow the power supplies to settle to their new set values during the first 10~ms. Each measurement cycle comprised injection and cooling followed by a range of 150 energies $E_\nu$ and was repeated for typically one hour. Adjacent energy ranges were chosen such that they mutually overlapped  by 50\%. After normalization to the electron current and the proxy ion current at $E_\mathrm{ref}$, the data from the various measurement cycles were combined into a relative recombination rate coefficient as function of electron-ion collision energy.

\begin{figure}
\includegraphics[width=\figurewidth]{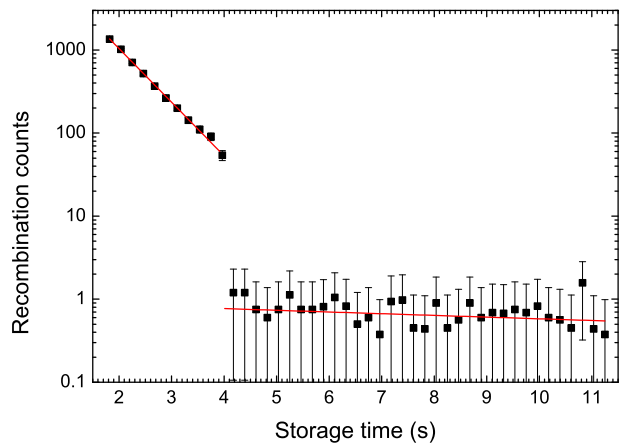}%
\caption{\label{fig:abs} (Color online) Number of recombination counts as function of storage time~$t$. The Cooler electron-beam was switched off at $t=4$~s and no Target beam was present at any time. The full line is a fit of exponentials to the data points.}
\end{figure}

The relative recombination rate-coefficient was put on an absolute scale by normalization to the recombination rate-coefficient $\alpha_0$ at zero eV electron-ion collision energy. The latter was separately measured by observing the storage lifetime of the ion beam \cite{Pedersen2005}. To this end the decay of the recombination count rate was monitored as a function of storage time (Fig.~\ref{fig:abs}). The lifetime of the stored ion beam in the storage ring is limited by  collisions with residual gas particles. When an electron beam is switched on, the lifetime is reduced even further by electron-ion recombination. The electron-ion recombination rate coefficient can thus be determined from the beam decay rate constants $\lambda^\mathrm{(on)}$ and $\lambda^\mathrm{(off)}$ measured with the Cooler switched on and off, respectively, i.e.,
 \begin{equation}\label{eq:alpha0}
   \alpha_0 = \frac{\lambda^\mathrm{(on)}-\lambda^\mathrm{(off)}}{n_eL/C}
 \end{equation}
where $n_e = 9.9\times10^6$~cm$^{-3}$ is the electron density, $L=1.5$~m is the length of the electron-ion interaction region, and $C=55.4$~m is the ring circumference. From exponential fits to the decay curves shown in Fig.~\ref{fig:abs} a value of $\alpha_0 = (5.3 \pm 0.2)\times 10^{-6}$~cm$^3$~s$^{-1}$ was obtained. Repeated measurements reproduced this value within the given fit error. Additional uncertainties arise from the electron density measurement and
the inaccurate knowledge of the interaction length. A deconvolution of the effects of the toroidal electron beam sections on the measured rate coefficient \cite{Lampert1996} was not carried out since the required knowledge of the recombination rate coefficient at higher energies is presently not available. The systematic uncertainty of the absolute rate coefficient scale is estimated to be 20\% at a 67\% confidence level \cite{Kilgus1992}.

The experimental energy spread is mainly determined by the velocity distribution of the Cooler electron beam which can be characterized by the longitudinal and transverse temperatures $k_BT_\|=0.15$~meV and $k_BT_\perp=10$~meV \cite{Kilgus1992}. These temperature values have been inferred from a previous experiment \cite{Schippers2001c} which was carried out under similar conditions as the present measurements. With these temperatures
the experimental energy spread is estimated \cite{Mueller1999c} to be 0.04~eV at an energy of 1~eV and 0.5~eV at 140~eV.

For the present measurements no dedicated effort has been made to calibrate the experimental energy scale more precisely than straight forwardly resulting from the merged-beams setup. The resulting  systematic uncertainty is particularly low at very low electron-ion collision energies \cite{Lestinsky2008a} and increases with increasing energy. A conservative estimate \cite{Kilgus1992} yields systematic uncertainties of 0.3 and 0.5~eV at electron-ion collision energies of 10 and 140 eV, respectively.

\section{Results and Discussion}\label{sec:res}

\subsection{Merged-beams recombination rate-coefficient}

\begin{figure}
\includegraphics[width=\figurewidth]{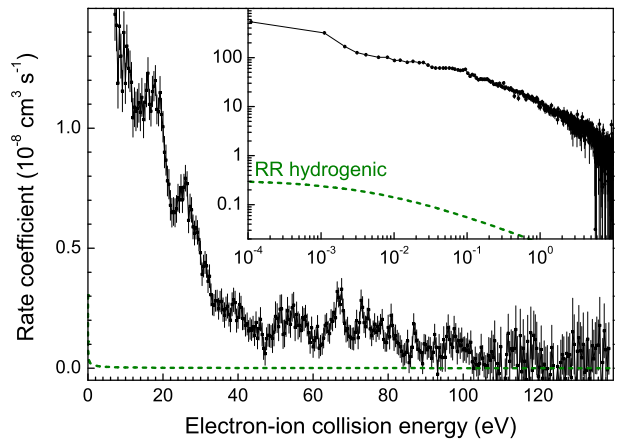}%
\caption{\label{fig:MB} (Color online) Measured merged-beams rate coefficient for electron-ion recombination of W$^{20+}$ ions as function of relative collision energy. The short-dashed curve is the calculated RR rate coefficient using a hydrogenic approximation \cite{Schippers2001c}. The inset shows the same data in a log-log representation and a finer energy binning emphasizing the rate coefficient at very low energies.}
\end{figure}

Figure \ref{fig:MB} shows the measured W$^{20+}$ merged-beams recombination rate coefficient as a function of electron-ion collision energy.
Most dramatically, the rate coefficient at energies at least up to 30~eV is characterized by a high level about three orders of magnitude above the RR rate coefficient estimated from a hydrogenic calculation. The measured rate coefficient decreases approximately monotonically from 0~eV up to an electron-ion collision energy of about 12 eV. From there on, broad resonance features become discernible up to the end of the experimental energy range. The widths of these features are much larger than the experimental energy spread. This indicates that the observed structures are most probably blends of many individually unresolved DR resonances.

The calculated RR rate coefficient  in Fig.~\ref{fig:MB} has been derived from a semi-classical hydrogenic formula for the RR cross section \cite{Schippers2001c} with nuclear charge $q=20$ and cutoff quantum number $n_\mathrm{max}=72$. This cutoff takes into account that recombined ions in loosely bound, high-$n$ W$^{19+}$([Kr]$4d^{10}\,4f^8\,nl$) Rydberg states are field ionized in the TSR dipole magnet before reaching the detector (see \cite{Schippers2001c} for details). Although the resulting merged-beams RR rate coefficient must be regarded as a crude estimate, it is quite clear that the contribution by RR to the measured rate coefficient is negligible below $\sim$100~eV.

The situation is quite similar to what has been observed in a single-pass merged-beams experiment with isoelectronic xenonlike Au$^{25+}$ ions \cite{Hoffknecht1998} where the measured rate coefficient also exceeded the RR rate coefficient by large factors. In the Au$^{25+}$ experiment, two well-separated broad groups of DR resonances were observed in the energy ranges of 15--40 eV and 65--85 eV, i.e., in nearly the same energy ranges where there are strong DR resonances in the present measurement. However the two Au$^{25+}$ resonance groups do not exhibit any substructure, whereas the present W$^{20+}$ DR spectrum has several distinct local maxima (Fig.~\ref{fig:MB}).

The inset of Fig.~\ref{fig:MB} enlarges the range of the lowest electron-ion collision energies where the most obvious feature is the rise of the rate coefficient at energies below 3~meV. This recombination rate enhancement is an artifact of the electron-ion merged-beams technique. It has been studied in detail previously \cite{Gwinner2000}. However, by its very limited energy range (0--2~meV in the present scan) this feature does not influence the plasma rate coefficient at the plasma temperatures where W$^{20+}$ is expected to be formed in fusion plasmas.

A detailed level assignment of the observed DR resonances is not possible because of the large number of excitation channels involved and because of the inherent uncertainties of atomic structure calculations for complex atomic systems. For the interpretation of the low-energy  DR spectrum in isoelectronic Au$^{25+}$\cite{Hoffknecht1998}, elaborate theoretical approaches \cite{Flambaum2002,Gribakin2003a} yielded some quantitative agreement with the experimental data but did not identify particular states.

\begin{figure}[t]
\includegraphics[width=\figurewidth]{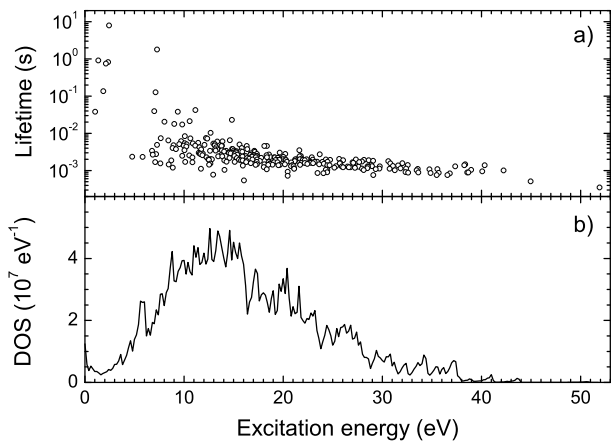}%
\caption{\label{fig:tau} a) Calculated excitation energies and lifetimes of 292 out of all 293 excited fine structure components of the W$^{20+}$([Kr]$4d^{10}\,4f^8$) ground configuration. The calculated lifetime of the $^3O_{12}$ state is off scale. b) Calculated density of states (DOS) comprising all doubly excited $4f^8\,nl$ states above the W$^{20+}$($4f^8\;^7F_6$) ground state with $n \leq 72$.}
\end{figure}

For the present work atomic-structure calculations have been carried out using Cowan's atomic structure code \cite{Cowan1981}. The W$^{20+}$ ground state was calculated to be the [Kr]$4d^{10}\,4f^8\;^7F_6$ state in agreement with previous theoretical results \cite{Kramida2009}. Figure \ref{fig:tau}a displays the lifetimes and excitation energies of all excited states belonging to the [Kr]$4d^{10}\,4f^8$ ground configuration. For a given state the lifetime was calculated from the M1 and E2 transition rates to all accessible energetically lower states. Transitions of higher order multipoles and configuration mixing could lead to a reduction of the calculated lifetimes, but were not considered here.

The majority of levels in the ground configuration have lifetimes less than 0.1~s and we expect most of the ion beam to relax towards the ground state.  However, from the present calculations it is found that there are 8 states with lifetimes larger than 0.1~s. The longest-lived excited states are the $^3O_{12}$, $^7F_0$, and $^5L_8$ states with lifetimes of 68000, 8.0 and 1.8~s, and excitation energies of 11.10, 2.43, and 7.276~eV, respectively. Since the waiting time after each injection was only 2~s in the present experiment, this suggests that not all stored ions were in the ground state when the recombination measurements took place.

In contrast, the situation in the single-pass Au$^{25+}$ experiment \cite{Hoffknecht1998} was quite different. There, the time between ion formation and recombination was only about 5~$\mu$s so that most probably a much larger fraction of ions  were in excited states when they underwent recombination in the electron target. Since DR resonance energies are different for different initial states, a large fraction of various metastable initial states could possibly explain the finding that the Au$^{25+}$ DR spectrum is less structured than the W$^{20+}$ spectrum. A more definitive statement about the population of metastable states in the ion beams would require an elaborate modeling of the radiation cascades involved in the de-excitation of the hundreds of excited states produced by the charge-stripping beam-foil interaction. Such a complex theoretical task is beyond the scope of the present paper.

Concluding from the statistical arguments above, the presently measured W$^{20+}$ DR resonance structure should for the most part be associated with excitations of the [Kr]$4d^{10}\,4f^8\;^7F_6$ ground state. Fine structure excitations to excited states of the ground configuration have excitation energies $E_\mathrm{exc}$ of up to about 52~eV (Fig.~\ref{fig:tau}a). Resonance energies of associated DR resonances can be estimated from the Rydberg formula for hydrogenic electron binding energies
\begin{equation}\label{eq:Ryd}
    E(n)=E_\mathrm{exc}-\mathcal{R}\frac{q^2}{n^2}
\end{equation} with
$\mathcal{R} \approx 13.606$~eV and $q=20$, which is reasonably accurate for sufficiently large principal quantum numbers $n$. Using $E_\mathrm{exc}=52$~eV one finds that $n=11$ is the smallest quantum number yielding a positive resonance energy. This is well below the cutoff quantum number $n_\mathrm{max}=72$. The smallest excitation energy of the W$^{20+}$ ground state which yields positive $E(n)$ for $n = 72$ is 1.05~eV, i.e., most of the 293 excited states of the W$^{20+}$ ground configuration contribute to the measured DR spectrum below 52~eV. The density of states (DOS) comprising all $(4f^8)_j\;nl$ states with $n\leq 72$ which are available for DR associated with fine structure excitations within the $4f^8$ shell can be easily calculated from the resonance energies (Eq.~\ref{eq:Ryd}) and statistical weights, i.e.\ $2n^2(2j+1)$, of the $(4f^8)_j\;nl$ Rydberg manifolds of DR resonances. As can be seen from Fig~\ref{fig:tau}b the resulting DOS is enormous, amounting to more than $10^7$ states per eV for energies in the 5--28~eV range.

Strong contributions to the observed DR resonance structure at energies above 52~eV can be expected from the dipole allowed core excitations $4d^{10}\,4f^8 \to 4d^9\,4f^9$, $4d^{10}\,4f^8 \to 4d^{10}\,4f^7\,5d$, and $4d^{10}\,4f^8 \to 4d^{10}\,4f^7\,5g$. According to the present single-configuration atomic-structure calculations the corresponding excitation energies are in the ranges 207--238~eV, 157--185~eV, and 291--318~eV, respectively. Using these excitation energies in equation~(\ref{eq:Ryd}) reveals that $4d^9\,4f^9\,nl$ resonances with $n=5$ or $6$ and $4d^{10}\,4f^7\,5d\,nl$ resonances with $n=6$ all have resonance energies in the 0--52~eV range and that resonances with higher Rydberg quantum numbers as well as all $4d^{10}\,4f^7\,5g\,nl$ resonances occur at energies above 52 eV.

It is questionable whether DR resonances associated with the just discussed dipole allowed core excitations can be held responsible for the steep increase of the experimental recombination rate coefficient below 47~eV where
it rises by almost 3 orders of magnitude from an average level of about $2\times 10^{-9}$~cm$^{3}$~s$^{-1}$ in the range 47--80~eV (Fig.~\ref{fig:MB}). Most probably, a large part of the huge low-energy rate coefficient is caused by DR associated with fine-structure core excitations within the [Kr]$4d^{10}\,4f^8$ ground configuration as represented in Fig.~\ref{fig:tau} and considered in the related discussion. Prominent DR resonances of this type have been previously identified in DR spectra of much simpler ions such as F-like Fe$^{17+}$ and O-like Fe$^{18+}$ \cite{Savin1997,Savin1999}.

\subsection{Plasma recombination rate-coefficient}

For the derivation of the W$^{20+}$ electron-ion recombination rate coefficient in a plasma the measured merged-beams rate coefficient (divided by the electron-ion relative velocity) has been convoluted with an isotropic Maxwellian energy distribution which is characterized by the plasma electron temperature $T_\mathrm{e}$.  In order to obtain an accurate result at temperatures far below 1~eV, the recombination rate enhancement mentioned above and the experimental electron energy spread have to be considered carefully  \cite{Schippers2001c}. However, these issues do not affect the plasma rate coefficient in the temperature range of interest for fusion plasmas and, therefore,  have been disregarded here.

\begin{figure}
\includegraphics[width=\figurewidth]{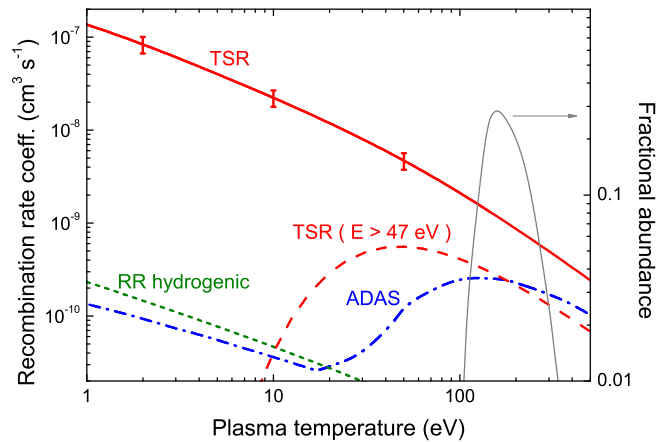}%
\caption{\label{fig:plasma} (Color online) Left scale: Rate coefficients for the recombination of W$^{20+}$ ions in a plasma. The thick full curve is the present experimentally derived result comprising RR and DR with resonance energies below 140~eV. Error bars denote the 20\% experimental systematic uncertainty.
The long-dashed curve results from the experimental data (Fig.~\ref{fig:MB}) with the collision energies restricted to the range 47--140~eV. The dash-dotted curve is the recombination rate coefficient from the ADAS data base \cite{adas,Whiteford2004}. The short-dashed curve is the result of a hydrogenic calculation \cite{Schippers2001c} for the RR rate coefficient. Right scale: The thin full line represents the calculated fractional abundance of W$^{20+}$ in a fusion plasma  \cite{Puetterich2005a}.}
\end{figure}

The experimentally derived W$^{20+}$ recombination rate in a plasma is shown in Fig.~\ref{fig:plasma}. It decreases nearly monotonically by more than two orders of magnitude across the displayed plasma temperature range. At all plasma temperatures the experimentally derived rate coefficient is very significantly larger than the W$^{20+}$ DR rate coefficient from the ADAS data base \cite{adas,Whiteford2004}. The difference becomes less drastic with increasing temperature. Moreover, it should be noted that the present result represents only  a lower limit because no data were taken at electron-ion collision energies higher than 140~eV. Nevertheless, the deviation of the experimental result from the theoretical prediction amounts to a factor of 4.3 at $kT_\mathrm{e}=160$~eV where the fractional abundance of W$^{20+}$ is predicted to peak in a fusion plasma \cite{Puetterich2005a} (Fig.~\ref{fig:plasma}).

Interestingly, the agreement between the ADAS plasma rate coefficient and the present result becomes considerably better at higher temperatures when all DR resonances below 47~eV (Fig.~\ref{fig:MB}) are excluded from the convolution procedure that yields the plasma rate coefficient. This finding suggests that the important low-energy DR resonances associated with fine-structure core excitations have been neglected in the calculation of the ADAS rate coefficient. Further support comes from the fact that at low temperatures  the ADAS data approach the RR rate coefficient (Fig.~\ref{fig:plasma}), which has been calculated using the hydrogenic cross section mentioned above, but with $n_\mathrm{max}=1000$ instead of $n_\mathrm{max}=72$.

\section{Summary and conclusions}\label{sec:conc}

Results from a storage-ring electron-ion recombination experiment with a tungsten ion have been presented. Because of low ion production rates and high loss rates during electron cooling of stored W$^{20+}$ ions, the experimental conditions have been less favorable as compared to previous experiments with lighter ions. Nevertheless, the numbers of mass-selected W$^{20+}$ ions stored and cooled were sufficient for performing meaningful DR measurements. The absolute electron-ion recombination rate coefficient, obtained with $\pm$20\% systematic uncertainty at electron-ion collision energies ranging from 0 to 140~eV. An additional error arising from the unknown population of metastable states in the ion beam cannot be estimated precisely, but does not change our finding that DR via fine-structure  core excitations is important.

The experimental recombination spectrum exhibits individually unresolved, huge DR resonances at very low electron-ion collision energies which strongly influence the plasma rate coefficient even at temperatures above 100~eV, where the fractional abundance of W$^{20+}$ is expected to peak in a fusion plasma. Because of the extraordinary complexity of the W$^{20+}$ atomic structure, no definitive assignment of the measured DR resonance features could be made. Atomic-structure calculations suggest that DR associated with fine-structure excitations of the W$^{20+}$([Kr]$4d^{10}\,4f^8$) ion core makes major contributions to the observed low-energy DR resonance strength. This fact seems to have been disregarded in the theoretical calculation of the W$^{20+}$ plasma DR rate coefficient which is used for plasma modeling by the nuclear fusion community. Their resulting rate coefficient is at least a factor of 4 lower than the present experimentally derived result.

We also find a factor of 4 difference between our results and the ad-hoc scaled W$^{20+}$ DR rate coefficient used in recent modeling of a tokamak plasma, for which the scaling factor was 0.97 \cite{Puetterich2008,Puetterich2010}. This large discrepancy shows that the derivation of rate coefficients from spectroscopic measurements on plasmas generally does not yield reliable results. Similar discrepancies should be expected for other charge states of tungsten. Consequently, the current modeling of the tungsten charge balance in fusion plasmas and of the associated radiative cooling bears large quantitative uncertainties which can only be reduced if further accurate experimental benchmarks for the relevant atomic cross sections and rate coefficients in the collision energy range below a few hundred eV become available.

\begin{acknowledgments}
We thank  the MPIK accelerator and TSR crews for their excellent support. MH, ON, and DWS were financially supported in part by the NASA Astronomy and
Physics Research and Analysis program and the NASA Solar Heliospheric Physics program. Financial support by the Max-Planck Society is gratefully acknowledged.
\end{acknowledgments}

\end{document}